\documentclass[onecolumn, appendixfloats, revtex4, apj, numberedappendix, iop, letterpaper]{openjournal}

\usepackage{orcidlink}
\usepackage{natbib}
\usepackage{amsmath}
\usepackage{needspace}
\usepackage{hyperref}

\def\pcyr{$\,$pc$\,$yr$^{-1}$}

\def\sqig{$\sim$}
\def\sun{$_\odot$}

\def\ctscm2s{cts\,cm$^{-2}$\,s$^{-1}$}

\def\ergscm2s{ergs\,cm$^{-2}$\,s$^{-1}$}
\def\ergss{ergs\,s$^{-1}$}

\journalinfo{}

\begin{document}

\shorttitle{Mundanity as an Explanation for the Fermi Paradox}
\shortauthors{R.\,H.\,D. Corbet }

\title{\vspace{-0.8cm}A Less Terrifying Universe? Mundanity as an Explanation for the Fermi Paradox\vspace{-1.5cm}
}

\author{Robin H.~D. Corbet\,\orcidlink{0000-0002-3396-651X}}
\affiliation{Center for Space Sciences and Technology, University of Maryland, Baltimore County, 1000 Hilltop Cir, Baltimore, MD 21250, USA; \href{corbet@umbc.edu}{corbet@umbc.edu}\\
X-ray Astrophysics Laboratory, Code 662 NASA Goddard Space Flight Center, Greenbelt Rd, Greenbelt, MD 20771, USA\\
%\affiliation{CRESST II}
Maryland Institute College of Art, 1300 W Mt Royal Ave, Baltimore, MD 21217, USA}
%\email{corbet@umbc.edu}
%\email{}

\begin{abstract}
Applying a principle of ``radical mundanity'', this paper examines explanations for the lack of strong evidence for the presence of technology-using extraterrestrial civilizations (ETCs) in the Galaxy - the Fermi paradox. With this principle, the prospect that the Galaxy contains a modest number of civilizations is preferred, where none have achieved technology levels sufficient to accomplish large-scale astro-engineering or lack the desire to do so. This consideration also leads to the expectation that no ETC will colonize a large fraction of the Galaxy, even using robotic probes, and that there are no long-duration high-power beacons. However, there is a reasonable chance that we may make contact on a short, by historical standards, timescale. This event would be momentous, but could still leave us slightly disappointed.  Such a Universe would be less terrifying than either of the two possibilities in the quote generally attributed to Arthur C. Clarke on whether we are alone or not. Also, if there is a modest number of ETCs in the Galaxy, that would suggest that there is a large number of planets with some form of life.
\end{abstract}


\maketitle

\Needspace*{7\baselineskip}
\section{Introduction} \label{sect:intro}

Science fiction author Arthur C. Clarke is purported to have said: ``Two possibilities exist: either we are alone in the Universe or we are not. Both are equally terrifying''.\footnote{The origin of this statement is unclear, and may well not be due to Clarke. See e.g. https://quoteinvestigator.com/2020/10/22/alone/}
Here the idea is explored that the truth may lie between Clarke's two terrifying alternatives in a rather more mundane, and so less terrifying, Universe.

``Modest'' assumptions fed into the Drake equation can predict a reasonable number of extraterrestrial civilizations (ETCs) within the Galaxy, and it has been proposed that it would take a relatively short time (astronomically speaking) for a civilization to spread across the entire Galaxy \citep[e.g.][]{Hart1975}.
It is expected that Galaxy-spanning civilizations would be seen via
technosignatures that might include: (i) detection of an artificial electromagnetic beacon; (ii) signs of astro-engineering such as waste heat from Dyson spheres  \citep{Stapledon1937, Dyson1960}; (iii) extraterrestrial artifacts on the Earth or other bodies in the solar system from previous visitations; (iv) direct evidence of extraterrestrials currently on the Earth itself.
However, so far there has not been a definite detection from any such search. The absence of detections from the Search for Extraterrestrial Intelligence (SETI) has been termed the ``Great Silence'' \citep[e.g.][]{Brin1983,Cirkovic2018}.
More generally, the lack of detection of electromagnetic signals or alien artifacts, given the predicted short time for an ETC to colonize the entire Galaxy, is termed the Fermi paradox \citep{Sagan1963,Shklovsky1966}.\footnote{But note that \citet{Gray2015} claims that it is neither a paradox nor Fermi's. In addition, \citet{Tarter2010} and \citet{Wright2018b} have argued that SETI searches have only explored a small region of parameter space.}

There have been a number of attempts to explain the Fermi paradox \citep[e.g.][]{Brin1983}, with several of these involving rather extreme explanations such as humanity being kept in a celestial zoo \citep[e.g.][and references therein]{Crawford2024} or that aliens have transcended to a form where they can no longer be recognized \citep[e.g.][]{Tsiolkovsky1932,Smart2012}.
On the other hand, some authors, such as \citet{Tipler1980}, consider the Great Silence as indicating that extraterrestrial civilizations do not exist. 

The goal of this paper is to consider not especially exciting solutions to the Fermi paradox, based on the concept that the most mundane explanation(s), if physically feasible, is/are most likely to be correct.
This follows from the Copernican mediocrity principle, where humanity does not occupy a special place in the Universe.

This paper first presents
a consideration of how technology levels can be quantified, and what the consequences of ETCs with high technology levels might be. This is followed by a mundane perspective on limits to technological development, and how this can impact both what technosignatures could be found, and how such limits may affect the motivations for an extraterrestrial technological civilization to explore the Galaxy for other such civilizations.
This paper then gives a brief selective review of searches for technosignatures, including some controversial claimed detections, and how these fit into the mundane view.

\section{Quantifying Technology Levels}

A common way that the SETI community discusses the technology level that could be achieved by an ETC is in terms of the Kardashev scale \citep{Kardashev1964}. A Type I civilization uses all the  energy available on a planet. A Type II civilization accesses all the energy emitted by a star, likely due to some variant of a Dyson sphere. Type III civilizations can control all energy emitted by a galaxy. 
\citet{Sagan1973} proposed that these should correspond to 10$^{23}$, 10$^{33}$, and 10$^{43}$ \ergss, respectively, with the Earth being approximately 0.72 on this scale.
For a Type II civilization, significant astro-engineering such as a Dyson sphere is likely required. It is less clear how a Type III civilization  could exist, except perhaps by constructing Dyson spheres around most stars in a galaxy. 

The Kardashev scale has limitations in that it only considers power utilization and no other facets of development. Other proposed scales include Sagan's, based on information storage \citep{Sagan1973}, and Barrow's, based on how small a scale objects can be manipulated \citep{Barrow1998}.
Other characterizations might involve 
the maximum spacecraft velocity achieved\footnote{The current human technology record holder is the Parker Solar probe, achieving 6.4$\times$10$^{-4}\,c$ at perihelion}
or
the intelligence level attained by computer systems. For example, a Type I civilization would have machines operating at the level of a single individual (artificial general intelligence or AGI), a Type II civilization would have machines operating at the level of a planet's total population, and a Type III civilization would have machines operating at the combined intelligence level of the populations of a combined galaxy. This is much harder to characterize given the difficulties involved in measuring intelligence \citep[e.g.][]{Hernandez2017}.
For a civilization where intelligence arises via stigmergy from a group \citep{Theraulaz1999}, rather than individuals, this is even harder to quantify.
Yet another measure of technological development might be extension of an organism's life span. But again, such a measure could well depend on the importance of individual organisms for a particular species. Thus, although the Kardashev levels are a crude measure, there is no obvious better replacement.

Clearly, what might be expected to be achieved by a civilization with extremely advanced technology is speculation. However, possibilities include undertaking astro-engineering projects such as constructing Dyson spheres, sending out swarms of interstellar robotic probes, or constructing powerful beacons that could be detected across the Galaxy.

\section{Colonizing the Galaxy}
\label{section:colonizing}
Even with current, or near-future, levels of technology, it may be possible for us to reach the nearest stars.
With present spacecraft speeds such as achieved by, for example, New Horizons\footnote{Approximately 2.8$\times$10$^{-5}\,c$ at Pluto}, it would take \sqig10$^5$ years to reach Proxima Centauri.
To reach higher speeds,
Project Daedalus \citep{Bond1978} and its offshoot Project Icarus \citep{Swinney2011} examined using nuclear-fusion powered rockets for interstellar travel, with the theoretical potential to reach 0.12\,$c$.
More recently, sending very small probes with solar sails powered by Earth-based lasers has been proposed for Breakthrough Starshot \citep{Parkin2018} which could then reach Proxima Centauri on a timescale of decades traveling at speeds of up to 0.2\,$c$.

Thus, even with current technology, travel to the nearest stars may well be possible, even if exceptionally challenging and extremely slow.
However, even if it is technically feasible to colonize the entire Galaxy, at least using self-replicating probes such as proposed by \citet{Hart1975}, for example, the colonizers will require a motivation to do this. The benefits obtained must outweigh the costs and potential risks.
Since it is unlikely that there would be much benefit from bringing physical objects back, the benefits must come from the scientific knowledge acquired and then sent back to the home planet(s). 

Apart from sending out probes to explore the Galaxy, another possible motivation for interstellar travel would be if the parent star of a civilization is reaching the end of its lifetime.
However, \citet{Gertz2020} argues that it is impossible for a civilization to migrate away from a dying star and that instead the civilization will endeavor to survive around the star at the end of its lifetime.
But even if a civilization located around a star at the end of its lifetime does need to emigrate to a different location, that does not necessarily lead to a broader expansion across the Galaxy.
Further, some white dwarfs may have a long-lived habitable zone \citep[e.g.][]{Zhan2024, Vanderburg2025, Barrientos2025}, and a civilization thus might be able to continue to exist around the remnant of its parent star. 

\Needspace*{7\baselineskip}
\section{A Mundane Perspective}

Here it is considered how a ``mundane'' perspective may reflect on the Fermi paradox. The main focus is on finding the least extreme situations that might explain this. The main premises are:
\begin{enumerate}
    \item There is a limit to technological development of a civilization, particularly in energy consumption. 
Extraterrestrial civilizations may be much more technologically advanced than contemporary Earth, but their technology does not include significant leaps equivalent to harnessing electricity, or rely on as yet unknown laws of physics. In this paper, this is termed a ``mundane'' technology level. 

\item There is a modest number of extraterrestrial civilizations that are at around this limit in the Milky Way Galaxy. Note that if our own civilization was unique in the Galaxy, this would imply that we are extraordinary, and so this would {\em not} be a mundane explanation and it would not be consistent with the Copernican mediocrity principle.

\end{enumerate}

The next sections consider the justification for these premises, and what their consequences would be.

\Needspace*{7\baselineskip}
\subsection{Fundamental Limits to Technological Development - Extraterrestrial Civilizations with Fairly Mundane Technology}

How far can technology continue to develop?
A lot of seminal papers on SETI were produced during the twentieth century during times of rapid scientific discovery and technological development \citep[e.g.][]{Mowery1999}. 
In modern science there are several major known unknowns. These include the nature of dark matter and dark energy, the asymmetry between matter and anti-matter, and how to unify quantum mechanics with general relativity. Given that dark energy, for example, was identified very recently, it can be expected that there is also an unknown number of unknown unknowns in science still remaining to be discovered. Nevertheless, it is unclear what the impacts of the resolution of these key scientific questions would be.
The scientific advances in understanding thermodynamics and electromagnetism had huge impacts on technology. Could any of the unknown unknowns have a comparable impact?
For example, if dark matter is determined to consist of WIMPs (Weakly Interacting Massive Particles) or axion-like particles as speculated, there is no clear way to incorporate that understanding into new technologies. 

Currently there is a lot of debate regarding the development of artificial intelligence (AI), even if such systems appear far from artificial {\em general} intelligence (AGI). While a technological singularity has been predicted by some \citep[e.g.][]{Vinge1993}, this may not occur. Or, if it does, it will affect the virtual world far more significantly than the physical world.

There are some indications that technological and engineering changes may already be becoming more modest.
\citet{Park2023} argued that papers and patents are becoming less ``disruptive'' over time.
Some have noted a slowing down of Moore's law \citep[e.g.][]{Lundstrom2022,Theis2017}.
Ramifications may be seen in the 
flattening of the growth rate of ``Top 500'' supercomputer list.\footnote{https://top500.org/statistics/perfdevel/}
Even if this is just a temporary slowdown in technological progress, a fundamental limit would eventually be reached.
Also, as possible additional constraints, \citet{Haqq2009} and \citet{Mullan2019} argue that sustainability places limits on the growth of civilizations,
\citet{Haqq2025} propose that stellar luminosity and mass provide a limit in the absence of very advanced technology,
and \citet{Balbi2025} consider constraints imposed by production of waste heat.

On the relatively near-term horizon, it should be possible to generate power via controlled nuclear fusion \citep[e.g.][]{Sadik2024}.
However, per capita energy usage in the United States has declined since 1979.\footnote{https://data.worldbank.org/indicator/EG.USE.PCAP.KG.OE?locations=US}
Thus, there may be no strong driver that compels a civilization with a fairly stable population level to move much further along the Kardashev scale.

\Needspace*{7\baselineskip}
\subsubsection{Implications of Technology Limits: Large-Scale Astro-Engineering is Impossible or Pointless or Both}

For the construction of large-scale astro-engineering projects that would be detectable by us, the extraterrestrial civilization requires both means and motive. The most commonly discussed such project would be the construction of a Dyson sphere or ring \citep{Niven1970}. These will not exist if either the incredible technology needed to construct one cannot exist, or if there is simply no need to have access to so much power. At lower technology levels, terra-forming a planet \citep[e.g.][]{Williamson1942,Sagan1961,Day2021} could be possible, but would not necessarily be obvious as an artifact to a distant observer.

\Needspace*{7\baselineskip}
\subsubsection{Implications of Technology Limits: No Long-Duration High-Power Beacons Co-located with a Civilization}

The case for a civilization maintaining a high-power signal to attempt to contact other civilizations for durations of time long enough for those civilizations to emerge is particularly challenging to justify. The transmitter would require substantial amounts of power for likely very long periods of time. This would make it very expensive, unless located in some place where the energy would not otherwise be useful for the civilization. If the duration of operation approaches the travel time across the Galaxy, then why not send probes instead of signals, since the probes can return other types of information?
An exception to this constraint could be if the beacon is not co-located with the civilization. In this case the power cost of the beacon would not be an issue as that power could not be used for other purposes.
Even in this case, it is not clear that there would be much motivation to operate a beacon for millions or billions of years.

\Needspace*{7\baselineskip}
\subsection{A Modest, Non-Zero, Number of Extraterrestrial Technological Civilizations}

The mundane view of how common technology using civilizations are in the Galaxy would be that they are neither extremely rare, nor extremely common.
This is consistent with the proposition by \citet{vonHoerner1961} as a basic assumption on other civilizations:
``Anything seemingly unique and peculiar to us is actually one out of many and is probably average''.
Most stars in the Galaxy are M type, which are \sqig10 times more common than G-type stars \citep[e.g.][]{Ledrew2001} such as our own G2V Sun. The mundane paradigm would tell us that these can't be hospitable for the development of technological civilizations, otherwise the Galaxy would likely be swarming with these. On the other hand, G-type stars are not that rare, if a technological civilization could only develop around early O stars then they would be extremely uncommon.

There are several M-type stars at reasonable distances from the Earth that are known to host planets including the nearest extra-solar star Proxima Centauri that has a planet (``b'') within the habitable zone \citep{Anglada2016}.  Barnard's Star, the fourth nearest star to the Sun, has at least four planets orbiting it \citep{Gonzalez2024, Basant2025}, although none lie in the habitable zone.
It is anticipated that we may be able to search for signs of life on nearby exoplanets in the relatively near future with missions such as the Habitable Worlds Observatory \citep[HWO, e.g.][]{Mamajek2024}. 
The principal problems for life around M stars are primarily intense stellar flares \citep[e.g.][]{France2016}, and the tidal locking for a planet in the habitable zone of its revolution and rotation periods \citep[e.g.][]{Barnes2017}.
Indeed, initial searches for an atmosphere on planets b and c around the M8V star TRAPPIST-1 with the James Webb Space Telescope have not been very promising \citep{Gillon2025}.

\Needspace*{7\baselineskip}
\subsubsection{Limits to Galactic Colonization, Even by Robots}

For significant colonization or exploration of the Galaxy by sending probes to almost all star systems would again have to have benefits that outweigh the costs. 
Or, in terms of Keynes' ``animal spirits'', greed must overcome fear (Keynes 1936).
Robotic exploration should be far cheaper, particularly if self-replicating 
von Neumann
probes \citep{Freitas1980}
are sent, where only a few initial probes need to be sent out.
The benefits depend on whether much is learned from each in situ exploration of a star system. Probably not much new scientific information would be provided by directly visiting, for example, many thousands of M stars, let alone billions. The cost would then be that of the initial launch, which potentially is not too great amortized over the entire lifetime of the civilization. However, there may be dangers in creating a vast number of (probably) intelligent devices. This is the ``Berserker'' hypothesis of deadly probes as described in the science fiction novels of Fred Saberhagen and discussed by, for example, \citet{Brin1983} and \citet{Cirkovic2009}. While the concept of Berserker probes may be extreme, the fear of this happening can be regarded as unexceptional. Indeed, this is something that concerns contemporary society with the development of AI - see for example \citet{Bostrom2014}, \citet{Amodei2016}, \citet{Joy2020}, and \citet{Muller2025}. 
In the mundane perspective, where other civilizations are not that much more advanced, a limit to exploration would arise when it was found that not much new was found from each encounter or other costs are incurred. In this case the benefit to risk ratio would decrease with each encounter and the passage of time.

\Needspace*{7\baselineskip}
\subsection{Life in a Mundane Galaxy, and Robustness}

In the proposed mundane Milky Way scenario, there will be a certain number of ETCs. As their technology develops they will reach a plateau in this development. For the Earth, the definition of the start of extensive advanced technology development is somewhat arbitrary, but could be taken as the start of the industrial revolution in around 1760 or the identification of radio waves in the 1880s.
If the typical lifetime of ETCs is $\gg$ the duration of technology use by humans to date, essentially all ETCs will be at this plateau technology level. This will be $\ll$ than Kardashev II, and there will be no ``magic'' technology such as faster-than-light travel.
Some fraction, perhaps all of them, will start to consider whether they are alone in the Galaxy, and commence some type of search, perhaps also starting with radio. They will likely initially encounter a Great Silence. Whether this persists as larger collecting area radio telescopes are constructed will depend on whether the typical distance between ETCs is much larger than the distance over which electromagnetic signals not deliberately constructed as beacons would be detected.
Then, they could potentially consider sending out robotic probes.

It must also be considered how robust the concept of mundanity is. That is, is it plausible that all ETCs, which may well have very different biologies and social systems all undergo the same limits and constraints?
The technological growth limits are based on the limits to which a knowledge of science allows a manipulation of the physical world. In this case if a limit does exist, it should apply to all ETCs, no matter their nature. The number of technological ETCs in the Galaxy is simply what it is. For a modest number of ETCs, the loosest constraints would be $\gg$ 1 and $\ll$ 10$^{11}$, and perhaps $\ll$ 10$^{10}$ if we exclude M-type stars. Somewhat arbitrarily taking the mean logarithmically would give \sqig 10$^{5}$ ETCs in the Galaxy.
The weakest link in the mundanity argument may be that we require a sufficient density of ETCs to halt interstellar exploration, or that all Galactic ETCs do not wish to send out self-replicating probes. Here the proposed primary limiting factor for this is that after encountering a certain number, $N_{ennui}$ of other ETCs are at the same technology plateau, the cost (in terms of the probes going ``berserk'' and the long time taken to find another ETC) is not worth the marginal gains from encountering additional ETCs. This effect could be compared to the concept of habituation where a creature's response to a repeated stimulus diminishes \citep[e.g.][]{Rankin2009}.
Also, encounters between different ETCs may occur before any self-replicating probes are constructed if plateau, or lower, technology can detect leakage radiation from neighboring ETCs. For example, the Square Kilometre Array \citep[SKA, e.g.][]{Dewdney2009} and the next-generation Very Large Array \citep[ngVLA, e.g.][]{Selina2018} are presently under construction and they will give major gains to the sensitivity of radio searches \citep{Siemion2015,Ng2022}. 
Indeed the SKA might be able to detect an Earth analog of distances up to 200pc \citep{Loeb2007} and planetary radar might be detected out to distances of kiloparsecs \citep{Sheikh2025}.
Even more sensitive radio telescopes beyond these are technologically possible, if expensive \citep[e.g.][]{Silk2025}.
If plateau-level technology is a bit boring to others already at that level, it means our pre-plateau technology civilization is of even less interest. Also, if life in general is rather commonplace, the biological ecosystem will be similar to that on many other planets. Hence, in this model reports of UFOs/UAPs are unlikely to be due to extraterrestrial visitors to the Earth. Thus the barriers to more wide-spread exploration by probes are fear of what the probes may become, and boredom in what they are finding.
The mundanity hypothesis can certainly be disproved by the discovery of an ETC's beacon, a Dyson sphere or other astro-engineering, or the presence in the solar system of an ETC artifact.

\subsection{Illustrative Estimates of Constrained Robotic Probe Exploration}

If detection of other plateau-level ETCs via leakage radiation is common, then it is expected that robotic exploration would be unusual. However, if that is not the case
then how would mundanity affect this exploration?

To obtain some rough estimates, a toy model is used which has a total of $N_{MW}$ technology plateau-level ETCs evenly distributed throughout the Milky Way's disc in an equilibrium such that ETC birth rate equals the death rate.
A simple model
of the Milky Way has radius, $r_{MW}$, of 17.5 kpc, and a height, $h_{MW}$, of 300 pc with ``hard edges'' and a uniform distribution of ETCs within this.
This model Milky Way then has a volume of \sqig2.9$\times$10$^{11}$ pc$^3$.

The density of ETCs is:
\[
\rho _{ETC} = \frac{N_{MW}}{(h_{MW} \pi r_{MW}^2)} 
\]

The constraint on this robotic exploration is the boredom/fear threshold. So, set the  interest level to decline to below the fear threshold at $N_{ennui}$ with direct contact.
The value of $N_{ennui}$ is, of course, a rather arbitrary choice but perhaps 100 could be reasonable.

So, if $N_{MW}$ = 100,000 and $N_{ennui}$ = 100, typically an ETC would explore 0.1\% of the Milky Way. 

For contact via probes, the first ``bubbles'' of ETCs aware of each other then arise when the first ETC in a region starts exploration.
Note that later there would not be independent contact bubbles in the mundane scenario, as they are expected to overlap in general. 

For simplicity, consider an ETC located in the center of the disc width (but not necessarily at the disc center).
The radius of the edge of the exploration front, $R_{exp}$ is the time $T$ during which exploration has been carried out multiplied by the probe expansion speed, $s_p$, where this includes the time required to replicate probes and launch capability.

For distances less than the thickness of the disc, the number of ETCs encountered, $N_{met}$, after time $T$ will then be:
\[
N_{met} \sim \rho _{ETC} \frac{4}{3} \pi (s_p T)^{3}
\]

Then, for an exploration rate in units of \pcyr:

\[
N_{met} \sim 1.45 \times10^{-11} N_{MW} (s_p T)^{3}
\]

For example, if $s_p = 0.001 c\, (\sim 3.1 \times 10^{-4}$ \pcyr), and N$_{MW}$ = 10$^5$, then after one million years the distance traveled will be 153 pc and N$_{met}$ = 5.2. This speed is a lot less than that considered by \citet{Hart1975} and \citet{Tipler1980} (0.1\,$c$) but is in line with considering ETCs with modest technology levels. However, it does require the ETCs to be more patient.

Taking N$_{met}$ = N$_{ennui}$, and rearranging:

\[
T = 4.1 \times10^3(\frac{N_{ennui} }{N_{MW}})^{1/3} s_p^{-1}
\]

For the same parameters as above, and N$_{ennui}$ = 100, T \sqig 1.3 Myr, and $d$ = 410 pc (which does take us out of the disk boundary).
Overall, the arguments for this taking a short time, compared to the age of the Milky Way, are similar to the arguments that the entire Galaxy could be colonized in a short time, it is simply made even shorter if there is no benefit to explore the entire Galaxy.

More generally, the volume of the explored region, $V_{exp}$, is given by the intersection of what would be the expansion sphere and the disk of the model Milky Way.
% Let the sphere have radius 
%R. Let the slab (thick disk of infinite lateral extent) have total thickness 
% h and be centred on the sphere center; let 
%a=h/2 be the half-thickness.
\[
V_{exp} =
\begin{cases}
\pi\!\left(R_{exp}^{2}h_{MW} - \dfrac{h_{MW}^{3}}{12}\right), & 0 \leq h_{MW} \leq 2R_{exp}, \\[8pt]
\dfrac{4}{3}\pi R_{exp}^{3}, & h_{MW} \geq 2R_{exp}.
\end{cases}
\]

\begin{figure}
\includegraphics[width=14cm,angle=270]{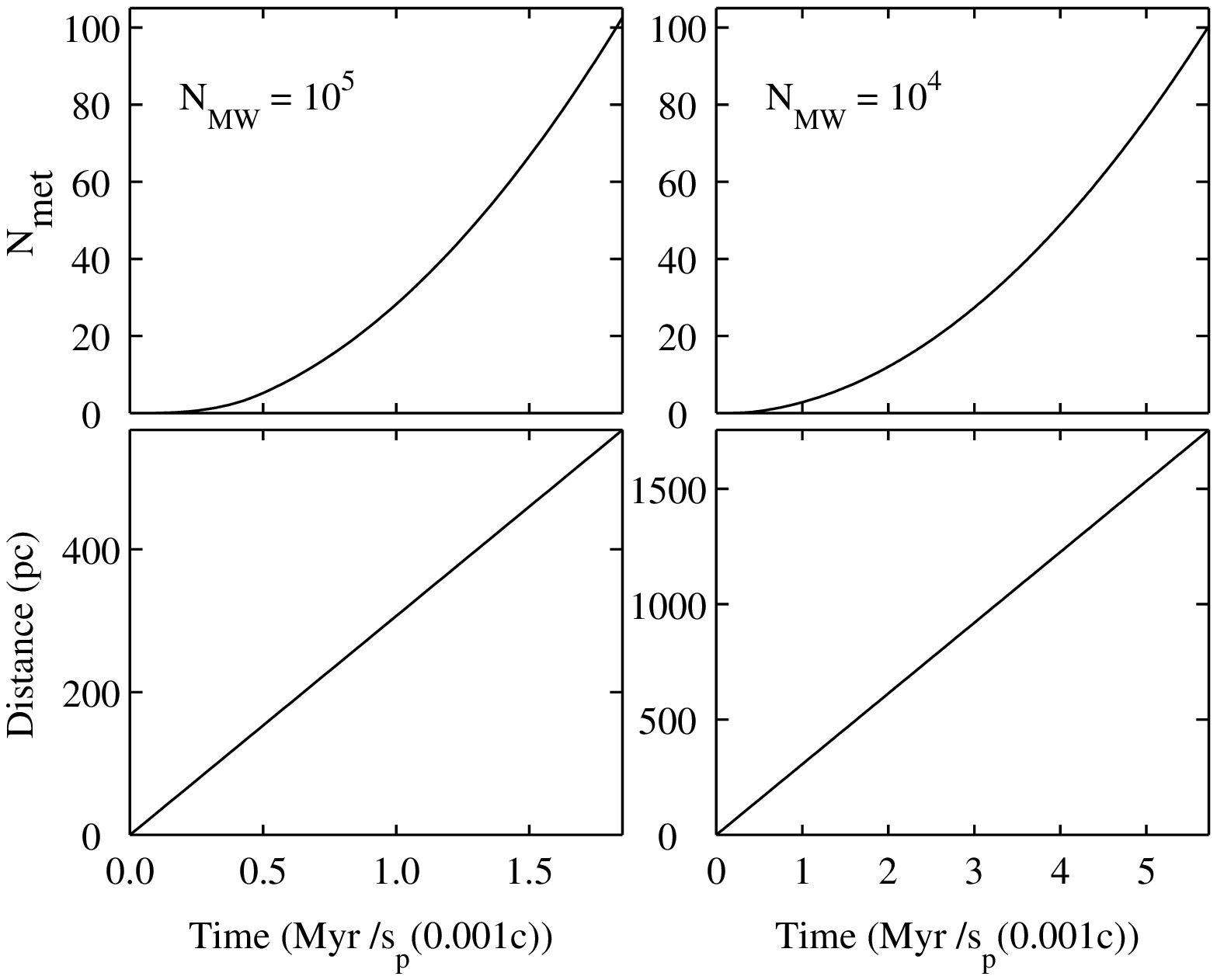}    
\caption{
The distance and number of ETCs encountered by exploration expanding at a rate, $s_p$ of 0.001\,$c$.
The left-hand figures are for $N_{MW} = 10^5$ and the right-hand figures are for $N_{MW} = 10^4$.
}
\label{fig:nmet_time}
\end{figure}

In Figure \ref{fig:nmet_time} are plotted the size of the exploration front, $R_{exp}$ and the number of ETC encounters, $N_{met}$, as a function of time for $N_{MW} = 10^{5}$ and $N_{MW} = 10^{4}$ for an exploration speed of $s_p = 0.001c$. The graphs remain the same shape for different expansion speeds, simply with a reduced or expanded timescale.
With this speed, in the $N_{MW} = 10^{5}$ case, 100 ETCs are encountered after $\sim1.85\,Myr/s_p(0.001c)$ and a distance of \sqig560\,pc, where $s_p(0.001c)$ is the expansion speed in units of 0.001$c$. For the $N_{MW} = 10^{4}$ case it takes $\sim5.7\,Myr/s_p(0.001c)$ and a distance of \sqig1.8\,kpc.

Note that more sophisticated models of robotic exploration of the Milky Way have been examined by, for example, \citet{Carroll2019} and lots of references therein.

\Needspace*{7\baselineskip}
\section{Technosignature Searches}

In this section, the current state of the search for ETCs is briefly and selectively reviewed, ranging from the detection of electromagnetic signals to physical objects, either at stellar distances or closer to home.
The results of these are then considered from a mundane point of view.

\Needspace*{7\baselineskip}
\subsection{SETI Across the Electromagnetic Spectrum}

Since \citet{Cocconi1959} proposed that radio waves are the best way to detect signals from ETCs, that waveband has dominated SETI searches. Prominent among these have been SERENDIP and SETI\@home surveys using the Arecibo Observatory \citep[e.g.][]{Werthimer2001},  Project META which surveyed the northern sky \citep{Horowitz1993}, and the contemporary Breakthrough Listen Initiative using the Green Bank Telescope and Parkes Observatory \citep{Isaacson2017,Worden2017}.

Even though radio has been the traditional wavelength for SETI, there has been a growing interest in using 
other wavelengths with optical searches in particular gaining in prominence.
This has been driven by the more tightly beamed signals that are possible at optical compared to radio wavelengths, and developments in laser technology which permit high-power brief signals to be generated \citep[e.g.][]{Tellis2015,Wright2018a}.
Generally such searches are designed to detect extremely brief monochromatic signals where a laser might momentarily outshine a star. These searches have been reviewed by \citet{Vukotic2021}.

At even shorter wavelengths, 
X-ray searches have also been considered by
\citet{Fabian1977}, \citet{Corbet1997}, and \citet{Lacki2025}.
In this case, the generation of such X-ray signals would likely require highly advanced astro-engineering, although the generation of an X-ray pulse through the explosion of nuclear weapons has also been proposed \citep{Elliot1973}. 

\Needspace*{7\baselineskip}
\subsection{Communication via Other Messengers}

Beyond the electromagnetic spectrum, neutrinos
have received some attention for either direct signaling \citep{Hippke2017,Jackson2020} or for timing communication \citep{Learned1994}. 
To date, gravitational waves have hardly been considered, as current technology can only detect merging black holes or neutron stars.
Beyond known messengers, if dark matter or dark energy are something that could be modulated and detected with appropriate technology, that is something that would presently be undetectable by us.

\Needspace*{7\baselineskip}
\subsection{Artifacts - Near to Far}

The scope for the detection of alien artifacts can range from the Earth itself, other bodies within the solar system, and at interstellar and stellar distances. \citet{Shostak2020} advocates for a greater emphasis on the search for artifacts rather than electromagnetic signals.
\citet{Stephenson1979} discussed the possibility of extraterrestrial cultures within our solar system, particularly the asteroid belt.
\citet{Benford2021}, for example, considered searches for alien artifacts on the Moon and other objects and proposed an equivalent of the Drake equation for artifacts.
To date there has been little in the way of direct searches for alien artifacts on the surface of the Moon or Mars, but see, for example, the discussion in \citet{Davies2013}.

\citet{Harris1986} describes a search for interstellar spacecraft based on the assumption that the power source would be either nuclear fusion or matter/antimatter annihilation, both of which would be expected to produce gamma-ray emission.

\Needspace*{7\baselineskip}
\section{Candidate Detections}

\Needspace*{7\baselineskip}
\subsection{Electromagnetic Signals}
Probably the most famous radio SETI candidate is the ``Wow'' signal found with the Ohio State Big Ear radio telescope \citep{Krauss1979}.
There has been no subsequent detection of a signal from the same location
\citep[e.g.][]{Gray1994, Harp2020,Perez2022}.
Recently a possible astrophysical explanation of this signal has been put forward by \citet{Mendez2024,Mendez2025} who
hypothesized that this signal was due to rapid brightening from stimulated emission of the hydrogen line caused by a strong transient radiation source, such as a magnetar flare or a soft gamma repeater.

Some other reports of candidate radio signals have been made from other searches, for example by \citet{Horowitz1993} and \citet{Painter2025} but none has been confirmed.

\Needspace*{7\baselineskip}
\subsection{Distant Artifacts}

At astronomical distances, for an object to be detected it needs to be sufficiently large and/or bright in some waveband, and Dyson spheres would match this in the Infrared.
\citet{Contardo2024} describe a recent search for Dyson spheres from an excess of infrared emission from main-sequence stars and reported 54 candidates. A similar search by \citet{Suazo2024} identified seven candidates.
These candidates were questioned, however, by \citet{Blain2024} who was also skeptical about the detectability of such objects. The deep irregular decreases in flux exhibited by F3 V star KIC 8462852, \citep[][sometimes referred to as ``Tabby's Star'' or other designations]{Boyajian2016}  have been postulated to be due to a partial Dyson sphere \citep[e.g.][]{Wright2016a}.
However, a number of astrophysical causes are also possible \citep[e.g.][]{Wright2016b,Bodman2016,Metzger2017}.
Radio \citep{Harp2016} and optical \citep{Schuetz2016} SETI observations of KIC 8462852 did not yield any  detection.

\Needspace*{7\baselineskip}
\subsection{Nearby Artifacts - the Solar System, and UFOs/UAPs on the Earth}

In our Solar System, perhaps the most dramatic, and controversial, claim for the detection of an alien artifact is the first interstellar object to be identified passing through the Solar System - `Oumuamua (1I/2017 U1).
\citet{Loeb2022} has proposed that `Oumuamua may have an artificial origin, 
perhaps as a fragment of a disintegrating Dyson sphere \citep{Loeb2023}. This has been disputed by, for example, \citet{Zhou2022} and \citet{Zuckerman2022}. Further, \citet{Davenport2025} caution that care must be taken when comparing interstellar bodies with solar system objects.

The Fermi paradox would of course be resolved if, in fact, intelligent aliens are already present or frequently visiting the Earth.
While there have been large numbers of reports of
Unidentified Flying Objects (UFOs), also referred to as Unidentified Anomalous Phenomena \citep[UAPs, or Unidentified Aerospace-Undersea Phenomena][]{Knuth2025}, to date there has been no strong evidence that any of these reports relate to alien visitation. Government reports on UFOs and UAPs include the U.S. Air Force's Projects Sign, Grudge and Blue Book, the Condon report \citep{Condon1968}, the Condign report\footnote{https://webarchive.nationalarchives.gov.uk/ukgwa/20121110115311/http://www.mod.uk/DefenceInternet/FreedomOfInformation/PublicationScheme/SearchPublicationScheme/UapInTheUkAirDefenceRegionExecutiveSummary.htm}
and the NASA UAP
Independent Study Team Report.\footnote{https://science.nasa.gov/uap/}
None of these reports found any definitive evidence for the current presence of active alien technology on the Earth.

\Needspace*{7\baselineskip}
\subsection{Technosignatures Misidentified as Natural Phenomena}

If technosignatures have been misidentified as natural phenomena, then the Fermi paradox could be resolved. This would probably be most likely for some phenomenon for which the astrophysics is so-far rather poorly understood, such as Fast Radio Bursts \citep[see e.g.][and citations therein]{Lorimer2024}
as discussed by \citet{Lingam2017}.

However, it has also been proposed that phenomena that appear to be well understood, may also be attributed to astro-engineering. For example, \citet{Vidal2019} discusses the use of pulsars for interstellar navigation, and whether or not millisecond pulsars might show signs of astro-engineering.
Similarly \citet{Learned2012}
suggests that the pulsations of Cepheid variables might be  modulated by an advanced civilization.

\Needspace*{7\baselineskip}
\section{A Mundane View of Candidate Detections}

In a mundane Galaxy we predict that there are no long-term powerful beacons and that only leakage radiation would be detected. This would not be likely to vary significantly. Thus, followup observations of a candidate should have a good chance of re-detecting it, but none of the candidate radio signals to date have passed this test.  

Construction of Dyson spheres and similar will not be possible  or desired. Therefore, no candidate Dyson sphere will be confirmed, and an astrophysical interpretation for KIC 8462852 is preferred.

With a reasonable population of ETCs in the Galaxy, the Earth is not likely to be a very interesting place to visit, and so all UFO/UAP reports will have causes other than visits by ETCs.
It may be somewhat less clear whether the mundane argument could be made against the claims of artificiality of `Oumuamua as it is unclear what would motivate such a flyby mission.

\Needspace*{7\baselineskip}
\section{What if One or Both Mundane Premises are False?}

The mundanity hypothesis is based on the premises that very highly advanced technology doesn't exist, and that there are a modest number of ETCs. What if one or both of these is/are not correct?

{\bf Many Civilizations, Very High Technology Level:} In this ``Star Trek'' scenario with high-technology levels throughout a crowded Galaxy, the Fermi paradox is indeed a paradox and an exotic mechanism is required to explain why there has been no detection of an ETC.

{\bf Very Few Civilizations, Very High Technology Level:} A solitary civilization would have a strong motivation to detect other ETCs. With ETCs being rare, each new one would be of great interest.
Possessing very advanced technology would enable it to create high-powered beacons, and spread robotic probes throughout the Galaxy. We then run into the Fermi paradox given our lack of the detection of probes. In this case there would finally be a breakdown of the Copernican mediocrity principle, as we would be one of a very small number of Technological Civilizations.

{\bf Many Civilizations, Mundane Technology Level:} In the most extreme form, every planet in its habitable zone would host an ETC. Thus Proxima Centauri b would be a promising location. While a candidate radio technosignature signal was once identified from Proxima Centauri, it was subsequently found to be an Earth-based artifact \citep{Sheikh2021}. 

{\bf Very Few Civilizations, Mundane Technology Level:} This situation naturally explains the Great Silence. However, it also invalidates one of the assumption behind the Fermi paradox, that a reasonable number of ETCs would arise and rather rapidly\footnote{i.e., on a timescale much less than the age of the Milky Way} spread throughout the Galaxy. Although it might be viewed as an extreme version of the mundanity paradigm, there would still be a drive for such civilizations to explore large volumes of the Galaxy. 
As noted in Section \ref{section:colonizing}, this is perhaps achievable by near-future technology on the Earth.

Thus, neither of these four scenarios appear as consistent with the inconclusive results of current SETI as does the double mundanity hypothesis.

\Needspace*{7\baselineskip}
\section{Conclusion}

The Fermi paradox may be explained if the Galaxy contains a modest number of technological civilizations, with technology levels that, while more advanced than contemporary Earth, are nowhere near the ``super-science'' levels that could result in readily detectable astro-engineering. The construction of powerful long-duration beacons would be unlikely, as would exploration of the entire Galaxy via robotic probes. With a modest number of technological civilizations at a modestly higher level of technology than ourselves, a detection of one of these via leakage radiation may not be too far off, historically speaking, with the SKA or a generation or two of radio telescopes beyond that. Although this would have profound implications in many ways, it may not lead to a huge gain in our technology level, and could leave us somewhat disappointed.
With a modest number of technological civilizations, this would imply that life in general would be rather common. We may thus hope that missions such as HWO \citep{Mamajek2024} or LIFE \citep{Quanz2022} have a reasonable chance of detecting this.
We note that a number of explanations for the Fermi paradox are based on extreme social or technology situations such as the zoo hypothesis, or aliens now being unrecognizable because they are so far advanced and have transcended to a different realm. Instead, the principle of mundanity, by definition, excludes such extreme situations. 

\section*{acknowledgements}
I thank Jacob Haqq-Misra, Ravi Kumar Kopparapu, Viktor Sandner and an anonymous referee for useful comments.
The work was supported in part by NASA under award number 80GSFC21M0002.

%\vspace{0.4cm}
\Needspace*{7\baselineskip}

%\bibliography{mundane_references.bib}
\end{document}